# *Declarative Semantics for Active Rules*

SERGIO FLESCA
*DEIS, Università della Calabria*
*87030 Rende, Italy*
*flesca@si.deis.unical.it*

SERGIO GRECO
*DEIS, Università della Calabria*
*and ISI-CNR*
*87030 Rende, Italy*
*greco@deis.unical.it*

**Abstract**

In this paper we analyze declarative deterministic and non-deterministic semantics for active rules. In particular we consider several (partial) stable model semantics, previously defined for deductive rules, such as well-founded, max deterministic, unique total stable model, total stable model, and maximal stable model semantics. The semantics of an active program $\mathcal{AP}$ is given by first rewriting it into a deductive program $\mathcal{LP}$, then computing a model $M$ defining the declarative semantics of $\mathcal{LP}$ and, finally, applying 'consistent' updates contained in $M$ to the source database. The framework we propose permits a natural integration of deductive and active rules and can also be applied to queries with function symbols or to queries over infinite databases.

## 1 Introduction

Active databases is an emerging technology combining techniques from databases, expert systems and artificial intelligence. The main peculiarity of this technology is the support for automatic 'triggering' of rules in response to events. Automatic triggering of rules can be useful in different areas such as integrity constraint maintenance, update of materialized views, knowledge bases and expert systems etc. (Widom & Ceri, 1996).

Active rules follow the so called *Event-Condition-Action* (ECA) paradigm; rules autonomously react to events occurring on the data, by evaluating a data dependent condition, and by executing a reaction whenever the condition is true. Active rules consist of three parts: *Event* (which causes the rule to be triggered), *Condition* (which is checked when the rule is triggered) and *Action* (which is executed when the rule is triggered and the condition is true). Thus, the semantics of a single active rule is that the rule reacts to a given event, tests a condition, and performs a given action. However, understanding the behavior of active rules, especially in the case of rules which interact with one another, is very difficult, and often the actions performed are not the expected ones. A very important issue in active databases is the development of tools which help in the design of programs with clear and intuitive semantics (Ceri & Fraternali, 1997).

The semantics of active rules are usually given in terms of execution models, which specify how and when rules will be applied. However, execution models are not completely satisfactory since their behavior is not always clear and could result in nonterminating computations. This is shown by the following example.

*Example 1*
Consider program $\mathcal{AP}$ where $+\mathtt{mgr}(\mathtt{X},\mathtt{P},\mathtt{D})$ (resp. $-\mathtt{mgr}(\mathtt{X},\mathtt{P},\mathtt{D})$) in the head of the rule means that, if the body is true, the atom $\mathtt{mgr}(\mathtt{X},\mathtt{P},\mathtt{D})$ is inserted into (resp. deleted from) the databases.

$$
\begin{array}{lll}
\mathtt{r}_1: & -\mathtt{mgr}(\mathtt{X},\mathtt{P},\mathtt{D}) & \leftarrow \quad -\mathtt{proj}(\mathtt{P}),\ \mathtt{mgr}(\mathtt{X},\mathtt{P},\mathtt{D})\cdot \\
\mathtt{r}_2: & +\mathtt{mgr}(\mathtt{X},\mathtt{P},\mathtt{D}) & \leftarrow \quad -\mathtt{mgr}(\mathtt{X},\mathtt{P},\mathtt{D}),\ \neg\mathtt{diff_mgr}(\mathtt{X},\mathtt{D})\cdot \\
\mathtt{r}_3: & -\mathtt{mgr}(\mathtt{X},\mathtt{P},\mathtt{D}) & \leftarrow \quad +\mathtt{mgr}(\mathtt{X},\mathtt{P},\mathtt{D}),\ \neg\mathtt{proj}(\mathtt{P})\cdot
\end{array}
$$

$$
\mathtt{diff_mgr}(\mathtt{X},\mathtt{D}) \quad \leftarrow \quad \mathtt{mgr}(\mathtt{X}',\mathtt{P},\mathtt{D}),\ \mathtt{X}' \neq \mathtt{X}\cdot
$$

Assume that the database $D$ consists of the tuples $\mathtt{proj}(\mathtt{p})$ and $\mathtt{mgr}(\mathtt{x},\mathtt{p},\mathtt{d})$ and that the initial update is $\delta = \{-\mathtt{proj}(\mathtt{p})\}$.[1] The meaning of the first rule is that if a project P is deleted (atom $-\mathtt{proj}(\mathtt{P})$ in the body of the rule) all managers of the project must be deleted. The second rule states that if a manager X of project P in the department D is deleted (atom $-\mathtt{mgr}(\mathtt{X},\mathtt{P},\mathtt{D})$ in the body) and there is not a second manager in the department D (literal $\neg\mathtt{diff_mgr}(\mathtt{X},\mathtt{D})$), then the manager X must be re-inserted (atom $+\mathtt{mgr}(\mathtt{X},\mathtt{P},\mathtt{D})$ in the head). The meaning of the third rule is that if a manager X of project P in the department D is inserted (atom $+\mathtt{mgr}(\mathtt{X},\mathtt{P},\mathtt{D})$ in the body) and the project P does not exist (literal $\neg\mathtt{proj}(\mathtt{P})$), then the manager X must be deleted (atom $-\mathtt{mgr}(\mathtt{X},\mathtt{P},\mathtt{D})$ in the head). Therefore, the rules define a sort of constraints on the insertion and deletion of tuples to guarantee the integrity of data.

The procedural evaluation of this active program applies first rule $r_1$ and then, alternately rules $r_2$ and $r_3$, which insert and delete the atom $\mathtt{mgr}(\mathtt{x},\mathtt{p},\mathtt{d})$ an infinite number of times. □

The evaluation of the active rules in the example above generates an infinite loop. Infinite loops can be avoided by a careful writing of rules. However, generally, it is very difficult to determine whether the procedural evaluation of a set of active results in a terminating computation. A second problem when dealing with active programs is that the future state of the database depends on the order in which active rules are fired — it seems that, under procedural semantics, the behavior of active rules is not easy to understand. Thus, two main properties have been identified for active rules: termination, which guarantees that the computation terminates in a finite number of steps, and confluence, which guarantees that the execution of rules always gives a unique outcome.[2] Several techniques have been designed for checking termination and confluence properties. These techniques are mainly based on the definition of 'compile-time' sufficient conditions. However, there are terminating and confluent programs which cannot be identified by these techniques.

[1] $\delta$ consists of a set of updates requested by users or transactions.
[2] A third important property is observable determinism. This property is not relevant in our framework since we consider only the insertion and deletion of tuples as actions.

Different solutions to these problems have been proposed in (Bidoit & Maabout, 1997), where a declarative semantics is associated to active rules, and in (Zaniolo, 1995; Lausen *et al.*, 1998), where active rules are modeled by means of deductive rules with an attribute denoting the state of the computation.

The solution proposed by Bidoit and Maabout (Bidoit & Maabout, 1997) is based on the computation of the well-founded semantics of a Datalog$^{\neg}$ program derived from the rewriting of the active program. The merits of this approach are that the semantics of programs is well-defined and the computation guarantees both termination and confluence in the general case.

However, the approach proposed in (Bidoit & Maabout, 1997) is not completely satisfactory since in many cases it does not capture the intuitive meaning of the active program.

*Example 2*
Consider the following program and the initial update $\delta = \{+\mathtt{confirm}(\mathtt{x},\mathtt{d})\}$.

$$+\mathtt{mgr}(\mathtt{X},\mathtt{D}) \leftarrow +\mathtt{confirm}(\mathtt{X},\mathtt{D})\cdot$$
$$-\mathtt{mgr}(\mathtt{X},\mathtt{D}) \leftarrow \mathtt{mgr}(\mathtt{X},\mathtt{D}),\ \neg+\mathtt{mgr}(\mathtt{X},\mathtt{D}),\ +\mathtt{confirm}(\mathtt{Y},\mathtt{D})\cdot$$

where the literal $\mathtt{mgr}(\mathtt{X},\mathtt{D})$ means that $\mathtt{X}$ is the manager of department $\mathtt{D}$ whereas the (update) literal $\neg+\mathtt{mgr}(\mathtt{X},\mathtt{D})$ means that the atom $\mathtt{mgr}(\mathtt{X},\mathtt{D})$ is not inserted in the database, i.e. the event associated with the insertion of the tuple does not happen. Since the atom $+\mathtt{mgr}(\mathtt{x},\mathtt{d})$ is inserted in the database, the second rule instantiated with $\mathtt{X}=\mathtt{x}$ and $\mathtt{D}=\mathtt{d}$ will not be fired; therefore the atom $\mathtt{mgr}(\mathtt{x},\mathtt{d})$ is not removed from the database. □

However, the semantics proposed in (Bidoit & Maabout, 1997) is not able to deduce this fact and it concludes that $\mathtt{mgr}(\mathtt{x},\mathtt{d})$ is undefined in the updated database. There are two problems with this semantics: the first is in the rewriting of active rules into deductive ones and the second is that the well-founded semantics in many cases is not satisfactory.

To overcome these problems we revise Bidoit and Maabout's technique by defining a different rewriting method and propose in addition to the well-founded, other declarative semantics.

The rest of the paper is as follows. In Section 2 we present preliminary definitions on active rules and Datalog. In Section 3 we present desirable properties which will be used to compare different semantics. In Section 4 we introduce deterministic and non-deterministic semantics for active rules. Then, in Section 5 we analyze the complexity of the semantics presented in the previous section. Finally, in Section 6 we compare declarative semantics, such as that proposed in this paper, with tools for checking termination and confluence, proposed for procedural semantics. We also discuss termination and confluence in the presence of functions symbols.

## 2 Preliminary definitions

In this section we introduce preliminary definitions and results on deductive and active rules. We also recall different versions of stable model semantics for DATALOG with negation.

### 2.1 Active and deductive rules

We assume finite countable sets of constants, variables and predicate symbols. A (simple) *term* is either a constant or a variable. A (standard) *atom* is of the form $p(t_1, \cdots, t_n)$ where $p$ is a predicate symbol and $t_1, \cdots, t_n$ are terms. A *literal* is an atom $A$ or its negation $\neg A$. Predicate symbols can be either *base* (EDB) or *derived* (IDB). Atoms and literals are either base or derived accordingly to their predicate symbols. An *update atom* (resp. literal) is of the form $+A$ or $-A$ where $A$ is a base atom (resp. literal).

A *rule* is of the form

$$A \leftarrow B_1, \cdots, B_n, C_1, \cdots, C_m$$

where $A$ is an atom, $B_1, \cdots, B_n$ are update literals and $C_1, \cdots, C_m$ are standard literals.

Rules can be either active or deductive. *Active rules* have an update atom in the head whereas *deductive rules* have a standard atom in the head. A deductive rule is called $\texttt{DATALOG}^{\neg}$ *rule* if the body does not contain update literals. A rule is said to be positive if it is negation free. Positive $\texttt{DATALOG}^{\neg}$ rules are called DATALOG rules.

Let $r$ be a rule; $H(r)$ and $B(r)$ represent, respectively, the head and the body of $r$. A ground DATALOG rule with no goals is called a *fact*. A *program* is a set of rules. A program $\mathcal{AP}$ is called *active* if it contains at least one active rule otherwise it is called *deductive*. A deductive program is called $\texttt{DATALOG}^{\neg}$ if does not contain update literals.

EDB predicate symbols form a relational database scheme $\mathcal{S}$, thus they are also seen as relation symbols. The set of all databases on a given schema $\mathcal{S}$ is denoted by $\mathbf{D}$. A database $D = \langle D^+, \overline{D} \rangle$ on $\mathcal{S}$ is a set of pairs of finite relations $\langle D^+(p), \overline{D}(p) \rangle$ on a countable domain $U$ (database domain), one for each $p$ in $\mathcal{S}$ such that $D^+(p) \cap \overline{D}(p) = \emptyset$. $D^+$ denotes the set of true facts in $D$ whereas $\overline{D}$ is the set of unknown facts in $D$. Moreover, we denote with $D^-$ the set of false facts in $D$, that is the set of facts which are not in the database. Observe that databases may contain, besides true facts, facts which are unknown. The reason for also considering unknown facts is to capture situations where the insertion or deletion of a tuple is undefined (for instance, under procedural evaluation, a tuple could be inserted and deleted indefinitely). A database $D$ on $\mathcal{S}$ is said to be *total* if $\overline{D} = \emptyset$, i.e. there are no unknown facts in $D$.

An *update program* $\mathcal{UP}$ is a pair $\langle \delta, \mathcal{AP} \rangle$, where $\delta$ is a set of update facts and $\mathcal{AP}$ is an active program. The set $\delta$ contains the *input update set* which is used to fire the active rules. An update program $\mathcal{UP}$ defines a mapping from $\mathbf{D}$ to $\mathbf{D}$. Thus, the application of an update program $\mathcal{UP}$ to a database $D_1$ on $\mathcal{S}$, denoted $\mathcal{UP}(D_1)$, gives a new database $D_2$ on $\mathcal{S}$. Our model is based on *durable changes*, first proposed in (Zaniolo, 1995), where the input update set $\delta$ is used to activate the rules in $\mathcal{AP}$ and only the updates derived from $\mathcal{AP}$ which will belong to the final update set are used to trigger again active rules. Thus, an update program $\mathcal{UP}$, on a database scheme $\mathcal{S}$, is a recursive (i.e., computable) function from $\mathbf{D}$ to $\mathbf{D}$. In addition to being computable, every update program $\mathcal{UP}$ is required to be *C-generic*, i.e. for any $D$ on $\mathcal{S}$ and for any isomorphism $\rho$ on $U - C$, $\mathcal{UP}(\rho(D)) = \rho(\mathcal{UP}(D))$, where the domain $C$ denotes the set of constants in the program. Informally speaking, the output of a program does not depend on the internal representation of the constants in the database not appearing in the program.

Given a database $D = \langle D^+, \overline{D} \rangle$ on $\mathcal{S}$ and an active program $\mathcal{AP}$, $\mathcal{AP}_D$ denotes the program obtained from $\mathcal{AP}$ by adding for each relation $p$ and for each tuple $t$ in $D^+(p)$ a fact $p(t)$ and for each tuple $t$ in $\overline{D}(p)$ a rule of the form $p(t) \leftarrow \neg p(t)$. [3] The set of constants in $\mathcal{AP}_D$ defines the *Herbrand universe* $H_{\mathcal{AP}_D}$ of $\mathcal{AP}_D$. The set of ground atoms built by using the constant in $H_{\mathcal{AP}_D}$ defines the *Herbrand base* $B_{\mathcal{AP}_D}$ of $\mathcal{AP}_D$ The *ground instantiation* of $\mathcal{AP}_D$ is denoted by $ground(\mathcal{AP}_D)$.

### *2.2 Semantics of* `DATALOG`$^\neg$

In this section we recall the definition of (partial) stable model semantics for `DATALOG`$^\neg$ programs. Partial stable semantics also applies to general logic programs, i.e. to programs where terms can be both simple or complex (a complex term is of the form $f(t_1, \cdots, t_n)$ where $f$ is a function symbol and $t_1, \cdots, t_n$ are terms).[4] We first recall the definition of partial stable model and next consider a restricted class of stable models called deterministic models.

#### *P-Stable Models*

Given a literal $A$, $\neg\neg A$ denotes $A$. Let $I$ be a set of ground literals; then $\neg I$ denotes the set $\{\neg A | A \in I\}$, and $I^+$ (resp., $I^-$) denotes the set of all literals (resp., negated atoms) in $I$. Given a `DATALOG`$^\neg$ program $\mathcal{LP}$ and a database $D$, we denote with $\overline{I} = B_{\mathcal{LP}_D} - (I^+ \cup \neg I^-)$ the set of facts in the Herbrand base which are undefined in the interpretation $I$. $I$ is a *(partial) interpretation* of $\mathcal{LP}_D$ if it is *consistent*, i.e. $I^+ \cap \neg I^- = \emptyset$. Moreover, if $I^+ \cup \neg I^- = B_{\mathcal{LP}_D}$, the interpretation $I$ is called *total*.

Letting $I$ be an interpretation for a program $\mathcal{LP}_D$, then the truth value of an atom $A \in B_{\mathcal{LP}_D}$ with respect to interpretation $I$, denoted by $I(A)$, is equal to (i) *true* if $A \in I$, (ii) *false* if $\neg A \in I$ and *undefined* otherwise, i.e. $A \in \overline{I}$. We assume the linear order $false < undefined < true$ and $\neg undefined = undefined$.

A rule $A \leftarrow A_1, \cdots, A_m$ in $ground(\mathcal{LP}_D)$ is satisfied w.r.t. an interpretation $I$

[3] A rule of the form $p(t) \leftarrow \neg p(t)$ states that $p(t)$ must be undefined since by assuming $p(t)$ false we derive $p(t)$ true and under the assumption that $p(t)$ is true $p(t)$ cannot be derived from the program since the body of the rule is false.

[4] For general logic programs we also assume a finite countable set of function symbols.

if $I(A) \geq min\{I(A_i) \,|\, 1 \leq i \leq m\}$. An interpretation $I$ is a model if all rules in $ground(\mathcal{LP})$ are satisfied. The semantics of logic programs is given in terms of partial stable model semantics (Gelfond & Lifschitz, 1988) which we briefly recall below.

An interpretation $M$ of $\mathcal{LP}_D$ is a *P-stable* (*partial stable*) model if it is the minimal model of the positive program $gl(\mathcal{LP}, M)$ obtained from $ground(\mathcal{LP})$ by replacing each negated body literal $\neg A$ with the complement of the truth value of $A$ w.r.t. $M$.

*Example 3*
Consider the program

```
a  ←  ¬b·
b  ←  ¬d·
c  ←  a, b·
```

This program has three P-stable models: $M_1 = \{\,\}$ (all atoms are undefined), $M_2 = \{a, \neg b, \neg c\}$ and $M_3 = \{\neg a, b, \neg c\}$. □

A P-stable model $M$ of $\mathcal{LP}_D$ is (i) *T-stable* (*total stable*) if it is a total interpretation of $\mathcal{LP}_D$, and (ii) *well-founded* if it is the intersection of all P-stable models of $\mathcal{LP}$. For instance, in the above example, $M_1$ is well-founded whereas $M_2$ and $M_3$ are T-stable. T-stable model was the first notion of stable model and was defined in (Gelfond & Lifschitz, 1988); the existence of a T-stable model for any program is not guaranteed. The well-founded model was introduced in (Van Gelder *et al.*, 1991) and is obviously unique. It is well known that a well-founded model exists for any program; therefore, the existence of at least one P-stable model is guaranteed as well.

Stable model semantics introduces a sort of non-determinism in the sense that programs may have more than one "intended" model (Gelfond & Lifschitz, 1988). A number of interesting subclasses of P-stable model have been recognized as possible "intended" models of a logic program.

In particular, the *M-stable* (*maximal stable*) models are those P-stable models that are not contained in any other P-stable model. Moreover, *L-stable* (*least undefined stable*) models are the M-stable models which leave a minimal set of elements of the Herbrand base undefined, and coincide with traditional total stable (*T-stable*) models whenever the set of undefined elements is empty.

*Example 4*
Consider the following program:

```
a·
b  ←  ¬c·
c  ←  ¬b·
p  ←  b, ¬p·
d  ←  a, ¬p, ¬e·
e  ←  a, ¬p, ¬d·
q  ←  ¬d, ¬q·
```

The P-stable models are: $M_1 = \{a\}$, $M_2 = \{a, b, \neg c\}$, $M_3 = \{a, \neg b, c, \neg p\}$, $M_4 = \{a, \neg b, c, \neg p, \neg d, e\}$, $M_5 = \{a, \neg b, c, \neg p, d, \neg e, \neg q\}$. $M_1$ is the well-founded model; $M_2$, $M_4$ and $M_5$ are the M- stable models; $M_5$ is also both L-stable and T-stable.

Now consider the above program without the first rule defining the predicate `a`. Then the P-stable models are: $M_1 = \{\neg a, \neg d, \neg e\}$, $M_2 = \{\neg a, \neg d, \neg e, b, \neg c\}$, $M_3 = \{\neg a, \neg d, \neg e, \neg b, c, \neg p\}$. $M_1$ is the well-founded model; $M_2$ and $M_3$ are the M-stable models; $M_3$ is also L-stable but not T-stable. (The atom $q$ is undefined in all P-stable models.) □

The *non-deterministic* version of the stable model semantics is based on the (non-deterministic) selection of a stable model (Greco *et al.*, 1995). Non-determinism offers a solution to overcome the limitations in expressive power of deterministic languages (Abiteboul & Vianu, 1991; Abiteboul *et al.*, 1990). For instance, it seems to be the only way to capture all polynomial-time queries without requiring the definition of an ordering on the domains, thus renouncing the data independence principle (Abiteboul *et al.*, 1994). The problem with stable model semantics is that the expressive power can blow up without control, so that polynomial-time resolution is no longer guaranteed. In fact, finding a T-stable model may require exponential time; worse, deciding whether a program has a T-stable model is $\mathcal{NP}$-complete (Marek & Truszczynski, 1991). Thus, it is possible that polynomial-time queries are computed in exponential-time, that is, it is possible to get untoward exponential time resolution.

## *Deterministic P-Stable Models*

In this section we present the concepts of deterministic P-stable models and of max-deterministic P-stable model. We point out that this issue is extensively analyzed in (Saccà & Zaniolo, 1997) also for the case of infinite universe and it is further investigated in relation to the notion of non-determinism. The notion of max-deterministic model was first introduced in (Saccà & Zaniolo, 1990) and its relevance was also discussed in (Greco & Saccà, 1997).

Let $\mathcal{LP}$ be a `DATALOG`$^\neg$ program, $D$ be a database on $\mathcal{S}$ and $M$ be a P-stable model of $\mathcal{LP}_D$. $M$ is *deterministic* if for every other P-stable model $N$ of $\mathcal{LP}_D$, $M$ and $N$ are not contradictory, i.e. $M \cup N$ is an interpretation. As proven in (Saccà & Zaniolo, 1997), two deterministic P-stable models are only an expression of assorted degrees of undefinedness inasmuch as there exists a P-stable model which includes both models.

The well-founded model is obviously a deterministic model — actually, as it is the intersection of all P-stable models, it is the minimum deterministic model. The family of deterministic models, denoted by $DM$, has an additional property: there exists a maximum element in the family, the *max-deterministic model*, which includes all other deterministic models and, therefore, can resolve all the differences among them.

*Fact 1*
*(Saccà & Zaniolo, 1997)* For every $\mathtt{DATALOG}^{\neg}$ program $\mathcal{LP}$, $\langle DM, \subseteq \rangle$ is a complete lattice such that the bottom is the well founded model and the top is the max-deterministic model. □

*Example 5*
Consider the following program $\mathcal{LP}$

```
a ← ¬b·
b ← ¬a·
c ← a·
c ← b·
c ← ¬d·
d ← ¬c·
```

The program $\mathcal{LP}$ has four P-stable models: $M_1 = \{\ \}$, $M_2 = \{c, \neg d\}$, $M_3 = \{a, \neg b, c, \neg d\}$ and $M_4 = \{\neg a, b, c, \neg d\}$. $M_1$ is well-founded, $M_3$ and $M_4$ are T-stable and $M_2$ is max-deterministic. □

Observe that deterministic P-stable models trade off minimal undefinedness to achieve determinism, but in different degrees. At the bottom, we find the well-founded model, which ensures better computability at the expense of more undefinedness. At the top, we find the max-deterministic model, whose clear semantic advantages are counterbalanced by computational drawbacks, as shown below.

## 3 Desirable properties of active programs

Several desirable properties for active rules have been introduced in the literature. The most important properties are termination and confluence.

*Termination* is the property which guarantees that the execution of active programs terminates in a finite number of steps. It is a crucial problem in active databases since rules may trigger each other recursively, and consequently, the non-termination of active rules execution is a constant threat.

*Confluence* is the property that the execution of active programs produces a final state (updated database) independent of the order of execution of (not prioritized) rules. In general, confluence is a desirable feature since the behavior of rules with a unique final state can be more easily understood and because most applications require a unique final state (e.g. materialized views applications).

In this paper we consider a different framework based on the use of declarative semantics. All the semantics considered guarantee termination of the update programs, since they are based on finite domains and stable model semantics. Furthermore, the first group of proposed semantics also guarantees confluence, since they identify a unique particular stable model as the intended model for the update programs. We also propose semantics that do not guarantee confluence, since they nondeterministically choose a model among a set of intended models.

Complexity and expressivity are also important interrelated properties. It is an open issue whether semantics with high complexity (and expressivity) are preferable. Here we consider declarative semantics which have low complexity and expressivity w.r.t. procedural semantics proposed in the literature.

For declarative semantics, several other desirable properties have been proposed (Bidoit & Maabout, 1997). A first requirement is that any semantics must be *founded*, that is, the set of (positive and undefined) atoms in the model must be derivable from the program. A second requirement is that the set of updates to be applied to a given database must be consistent.

*Definition 1*
*A set of update literals $M$ is said to be*

1. conflict free *if it does not contain two update facts of the form $+A$ and $-A$;*
2. consistent [5] *if for each atom $+A \in M$ (resp. $-A \in M$) it must be the case that $\neg - A \in M$ (resp. $\neg + A \in M$).* □

Finally, we consider a property of knowledge ordering among databases (i.e. set of atoms which can be either true or undefined).

*Definition 2*
*Given two databases $D_1$ and $D_2$, we say that $D_2$ is* more informative *than $D_1$ or, equivalently, that $D_1$ is not more informative than $D_2$ ($D_1 \preceq D_2$) if $\overline{D}_2 \subseteq \overline{D}_1$.* □

This property is important when we are interested in databases having maximal sets of true and false facts, and, therefore, our transformations should minimize the set of unknown facts.

## 4 Declarative Semantics for Update Programs

As discussed in the Introduction, the semantics proposed in (Bidoit & Maabout, 1997) is not able to capture the intuitive meaning of all programs. Thus, in this paper we propose different semantics for active rules. More specifically, we consider a different rewriting of active rules into deductive ones and analyze several (partial) stable model semantics. Let us start by introducing how sets of updates are applied to databases.

*Definition 3*
*Let $D = \langle D^+, \overline{D} \rangle$ be a database and let $M$ be a consistent set of update literals. Then, the application of $M$ to $D$, denoted $M(D)$, gives a new database $D_1$ defined as follows:*

1. *$p \in D_1^+$ if $+p \in M$ or ($p \in D^+$ and $\neg - p \in M$);*
2. *$p \in \overline{D}_1$ if one of the following conditions is true*
   (a) *$p \in \overline{D}$ and $+p, -p \notin M^+$;*
   (b) *$p \in D^+$ and $-p \in \overline{M}$;*

[5] This property has been called *well founded* in (Bidoit & Maabout, 1997).

*Algorithm 1*
Rewrite update programs.
**Input** Update program $\mathcal{UP} = \langle \delta, \mathcal{AP} \rangle$;
**Output** DATALOG$^\neg$ program $st(\mathcal{UP})$;
**Method**

1. insert into $st(\mathcal{UP})$ the deductive rules in $\mathcal{AP}$;
2. For each active rule $+a(t) \leftarrow B$ (resp., $-a(t) \leftarrow B$) in $\mathcal{AP}$ insert into $st(\mathcal{UP})$ the rule $+a(t) \leftarrow B, \neg ck_a(t)$ (resp., $-a(t) \leftarrow B, \neg ck_a(t)$) where $ck_a$ is a new predicate symbol;
3. For each rule $r$ in $st(\mathcal{UP})$ replace each update atom $+p(t)$ (resp. $-p(t)$) in the body with $+p(t) \vee \ p'(t)$ (resp., $-p(t) \vee p''(t)$), where $p'$ and $p''$ are new IDB predicate symbols;
4. For each new predicate symbol $ck_a$ with arity $n$, introduced in Step 2, add a rule $ck_a(X_1, \cdots, X_n) \leftarrow +a(X_1, \cdots, X_n), -a(X_1, \cdots, X_n)$;
5. For each update atom $+p(t)$ (resp. $-p(t)$) $\in \delta$ insert into $st(\mathcal{UP})$ the fact $p'(t)$ (resp., $p''(t)$) (introduced in the body of rules in Step 3);
6. Rewrite the program in clausal form by eliminating disjunctions;
7. Return $st(\mathcal{UP})$ where every update atom $+a(t)$ (resp., $-a(t)$) is interpreted as a standard atom, i.e., $+a$ and $-a$ are new standard predicate symbols. □

Fig. 1. Rewriting of update programs

(c) *$p \in D^-$ and $+p \in \overline{M}$.* □

Thus, an atom $p$ is undefined in the new database $D_1$ if (1) it was undefined in the old database $D$ and there is no evidence about its insertion or deletion, or (2) it was true (resp. false) in $D$ and its deletion (resp. insertion) is undefined in $M$.

The algorithm reported in Figure 1 rewrites an update program $\mathcal{UP}$ into a deductive program denoted as $st(\mathcal{UP})$.

Essentially, the algorithm rewrites update atoms taking into account information contained in the input update $\delta$, and adds 'temporary' rules which are used to guarantee the consistency of the model, i.e. any model containing an atom of the form $+a(t)$ must also contain the literal $\neg - a(t)$. In particular, for each update $+a(t)$ (resp. $-a(t)$) in $\delta$ a fact $a'(t)$ (resp. $a''(t)$) is added to the program (Step 5) and every update atom $+a(u)$ (resp. $-a(u)$) in the body of a rule is replaced by a disjunction $+a(u) \vee a'(u)$ (resp., $-a(u) \vee a''(u)$) which takes into account that the insertion (resp. deletion) of an atom can be derived from the program or it is contained in the input update set $\delta$ (Step 3). Moreover, programs are subsequently rewritten in standard form by eliminating disjunctions.

The derived program contains update literals which are interpreted as standard ones, i.e. an update atom $+p(t)$ will be interpreted, in the rewritten program, as a standard atom with predicate symbol $+p$. The declarative semantics of the rewritten deductive program $st(\mathcal{UP})$ gives a set of literals also containing elements with predicate symbols of the form $+p$ and $-p$; these atoms will be interpreted as insertions and deletions to be applied to the source database. The following example clarifies the rewriting of update programs.

*Example 6*
Consider the update program $\mathcal{UP} = \langle \delta, \mathcal{AP} \rangle$ of Example 1 where $\delta = \{-\texttt{proj(p)}\}$ and $\mathcal{AP}$ consists of the following rules

$$
\begin{array}{lll}
-\texttt{mgr(X,P,D)} & \leftarrow & -\texttt{proj(P)},\ \texttt{mgr(X,P,D)}\cdot \\
+\texttt{mgr(X,P,D)} & \leftarrow & -\texttt{mgr(X,P,D)},\ \neg\texttt{diff_mgr(X,D)}\cdot \\
-\texttt{mgr(X,P,D)} & \leftarrow & +\texttt{mgr(X,P,D)},\ \neg\texttt{proj(P)}\cdot \\
\texttt{diff_mgr(X,D)} & \leftarrow & \texttt{mgr(X}',\texttt{P,D)},\ \texttt{X}' \neq \texttt{X}\cdot
\end{array}
$$

The deductive program derived from the application of Algorithm 1 before rewriting the disjunctions (Step 6) is

$$
\begin{array}{lll}
-\texttt{mgr(X,P,D)} & \leftarrow & (-\texttt{proj(P)} \vee \texttt{proj}''\texttt{(P)}), \\
 & & \texttt{mgr(X,P,D)},\ \neg\texttt{ck_mgr(X,P,D)}\cdot \\
+\texttt{mgr(X,P,D)} & \leftarrow & (-\texttt{mgr(X,P,D)} \vee \texttt{mgr}''\texttt{(X,P,D)}), \\
 & & \neg\texttt{diff_mgr(X,D)},\ \neg\texttt{ck_mgr(X,P,D)}\cdot \\
-\texttt{mgr(X,P,D)} & \leftarrow & (+\texttt{mgr(X,P,D)} \vee \texttt{mgr}'\texttt{(X,P,D)}), \\
 & & \neg\texttt{proj(P)}\ \neg\texttt{ck_mgr(X,P,D)}\cdot \\
\texttt{proj(p)}''\cdot & & \\
\texttt{ck_mgr(X,P,D)} & \leftarrow & +\texttt{ck_mgr(X,P,D)},\ -\texttt{ck_mgr(X,P,D)}\cdot \\
\texttt{diff_mgr(X,D)} & \leftarrow & \texttt{mgr(X}',\texttt{P,D)},\ \texttt{X}' \neq \texttt{X}\cdot
\end{array}
$$

The final program is obtained by replacing rules with disjunctions in their bodies by standard rules. For instance, the second rule defining the predicate $+\texttt{mgr(X,P,D)}$ is replaced by the rules

$$
\begin{array}{lll}
+\texttt{mgr(X,P,D)} & \leftarrow & \texttt{mgr}^{-}\texttt{(X,P,D)},\ \neg\texttt{diff_mgr(X,D)},\ \neg\texttt{ck_mgr(X,P,D)}\cdot \\
\texttt{mgr}^{-}\texttt{(X,P,D)} & \leftarrow & -\texttt{mgr(X,P,D)}\cdot \\
\texttt{mgr}^{-}\texttt{(X,P,D)} & \leftarrow & \texttt{mgr}''\texttt{(X,P,D)}\cdot
\end{array}
$$

□

In the following we will consider only semantics which associate to any update program a set of stable models. Moreover, if the set of stable models is empty we reject the program whereas if it contains more than one element we select, non deterministically, one model. However, if the set of stable models is a singleton, we say that the semantics is *deterministic* whereas if all models agree on the database atoms we say that the semantics is *confluent*. It is obvious that deterministic semantics are also confluent (for instance, the well-founded semantics is an example of deterministic semantics).

Consider an update program $\mathcal{UP}$ and a database $D$. Then, we denote with $\mathcal{XS}$ a generic stable model semantics and with $\mathit{SEM}_{\mathcal{XS}}[\mathcal{UP}, D]$ the model of $st(\mathcal{UP})_D$ under the $\mathcal{XS}$ semantics. Moreover, given a set of literals $M$, we denote with $\hat{M}$ the set of update literals in $M$. The following definition introduces the application of an update program $\mathcal{UP}$ to a database $D$.

*Definition 4*
Let $\mathcal{UP} = \langle \delta, \mathcal{AP} \rangle$ be an update program and let $D$ be a database. Then, the application of $\mathcal{UP}$ to $D$ under stable model semantics $\mathcal{XS}$ denoted $\mathcal{UP}_{\mathcal{XS}}(D)$, is defined as $\Gamma_{\mathcal{XS}}(\delta(D))$ where $\Gamma_{\mathcal{XS}} = \widehat{\mathit{SEM}}_{\mathcal{XS}}[\mathcal{UP}, D]$. □

The application of $\mathcal{UP} = \langle \delta, \mathcal{AP} \rangle$ to $D$ is carried out in three steps: (i) first the set of update atoms which are derived from the standard version of $\mathcal{UP}$ applied to $D$ is computed, (ii) next, $\delta$ is applied to $D$ and, finally, (iii) the set of derived update atoms is applied to the current database. Observe that in the application of an update program $\mathcal{UP}$ to a database $D$, we first apply the input update set $\delta$ and next the updates derived from the active program $\mathcal{AP}$. Thus, we give priority to the active program since it defines a sort of constraints which must be satisfied — updates in $\delta$ could be nullified by the rules in the program.

Observe that our semantics is based on *durable changes*, first proposed in (Zaniolo, 1995). Durable changes are updates which will belong to the final update set. Thus, given an update program $\mathcal{UP} = \langle \delta, \mathcal{AP} \rangle$ and a database $D$, the meaning of a negated update literal $\neg + A$ appearing in $\mathcal{AP}$, is that the atom $A$ is not inserted in the database by the application of $\mathcal{UP}$ to $D$. This, implies that rules are only triggered by updates which will belong to the final update set.

*Proposition 1*
Let $\mathcal{UP}$ be an update program and let $D$ be a database. Then, $\hat{SEM}_{\mathcal{XS}}[\mathcal{UP}, D]$ is consistent for every stable model semantics $\mathcal{XS}$.

**Proof** The proof derives directly from the rewriting of programs. Let $M$ be the unique model for $st(\mathcal{UP})_D$ under semantics $\mathcal{XS}$. Each update atom $+a(t)$ is in $M$ only if $\neg ck_a(t) \in M^-$. Moreover, if $+a(t)$ is true and $ck_a(t)$ is false w.r.t. $M$, then $-a(t)$ must be false. □

*Corollary 1*
Let $\mathcal{UP}$ be an update program and let $D$ be a database. Then, for each stable model semantics $\mathcal{XS}$, $\hat{SEM}_{\mathcal{XS}}[\mathcal{UP}, D]$ is conflict-free.

**Proof *(Sketch).*** Straightforward. □

We next analyze the impact of several deterministic and non-deterministic semantics w.r.t. the properties introduced in Section 3 and analyze their complexity and expressivity. In the next subsection we analyze confluent semantics whereas in subsection 4.2 we analyze non-confluent semantics based on stable models.

### *4.1 Confluent semantics*

Confluence is the property that given a database, the execution of active programs produces a new database independent of the order of execution of rules. The following example shows that the procedural evaluation of active rules may give alternative results which are not all intuitive

*Example 7*
Assume a database consisting of the four `emp`, `dept`, `mgr` and `prom` where i) a tuple `emp(e,d)` means that employee `e` works for the department `d`, ii) a tuple `dept(d)` means that `d` is a department iii) a tuple `mgr(e,d)` means that `e` is a manager of department `d`, and iv) a tuple `prom(e,d)` means that employee `e` is a promoted manager of department `d`. Consider the following program.

$$\begin{array}{llcl} \mathtt{r}_1: & -\mathtt{emp}(\mathtt{X},\mathtt{D}') & \leftarrow & +\mathtt{prom}(\mathtt{X},\mathtt{D}),\ \mathtt{emp}(\mathtt{X},\mathtt{D}')\cdot \\ \mathtt{r}_2: & +\mathtt{mgr}(\mathtt{X},\mathtt{D}) & \leftarrow & +\mathtt{prom}(\mathtt{X},\mathtt{D})\cdot \\ \mathtt{r}_3: & -\mathtt{dept}(\mathtt{D}) & \leftarrow & -\mathtt{emp}(\mathtt{X},\mathtt{D}),\ \neg\mathtt{diff}(\mathtt{E},\mathtt{D})\cdot \\ \mathtt{r}_4: & -\mathtt{emp}(\mathtt{X},\mathtt{D}) & \leftarrow & -\mathtt{dept}(\mathtt{D}),\ \mathtt{emp}(\mathtt{X},\mathtt{D})\cdot \\ \mathtt{r}_5: & -\mathtt{mgr}(\mathtt{X},\mathtt{D}) & \leftarrow & -\mathtt{dept}(\mathtt{D}),\ \mathtt{mgr}(\mathtt{X},\mathtt{D})\cdot \end{array}$$

$$\mathtt{diff}(\mathtt{X},\mathtt{D}) \leftarrow \mathtt{emp}(\mathtt{X}',\mathtt{D}), \mathtt{X}' \neq \mathtt{X}\cdot$$

$$\mathtt{diff}(\mathtt{X},\mathtt{D}) \leftarrow \mathtt{mgr}(\mathtt{X}',\mathtt{D}), \mathtt{X}' \neq \mathtt{X}\cdot$$

The first two rules say that if an employee $\mathtt{X}$ working for department $\mathtt{D}'$ is promoted to be manager of department $\mathtt{D}$, then the tuple $\mathtt{emp}(\mathtt{X},\mathtt{D}')$ must be deleted and a tuple $\mathtt{mgr}(\mathtt{X},\mathtt{D})$ must be inserted into the database. The third rule says that if an employee $\mathtt{X}$ working for the department $\mathtt{D}$ is deleted and there is no other employee working for $\mathtt{D}$, then department $\mathtt{D}$ must be deleted as well. The last two active rules say that if a department $\mathtt{D}$ is deleted, all employees and managers of $\mathtt{D}$ must be deleted too.

Assume now that the initial database is $D = \{\mathtt{dept}(\mathtt{d1}), \mathtt{emp}(\mathtt{e1},\mathtt{d1})\}$ and that the input update set is $\delta = \{+\mathtt{prom}(\mathtt{e1},\mathtt{d1})\}$, i.e. the employee $\mathtt{e1}$ is promoted to be a manager of department $\mathtt{d1}$. The procedural evaluation of the active rules gives several outcomes depending on the sequence of rules evaluation. Three possible instances are the following: $D_1 = \{\mathtt{dept}(\mathtt{d1}), \mathtt{mgr}(\mathtt{e1},\mathtt{d1}), \mathtt{prom}(\mathtt{e1},\mathtt{d1})\}$ (derived from the execution sequence $(r_1, r_2, r_3)$ $D_2 = \{\mathtt{mgr}(\mathtt{e1},\mathtt{d1}), \mathtt{prom}(\mathtt{e1},\mathtt{d1})\}$ and (derived from the execution sequence $(r_1, r_3, r_5, r_4, r_2)$ and $D_3 = \{\mathtt{prom}(\mathtt{e1},\mathtt{d1})\}$ (derived from the execution sequence $(r_1, r_3, r_2, r_5, r_4)$). □

In general, deterministic semantics guarantees that the rewritten deductive programs have a unique minimal model and, therefore, that the final state of the database is unique and well-defined.

In this subsection we consider several confluent semantics for update programs. We first analyze the standard well-founded semantics, the max deterministic semantics. Then we consider the restrictions of well-founded and max-deterministic semantics to total transformations, i.e. to transformations which do not introduce undefined tuples into the database. Finally, we analyze the unique stable model semantics.

#### *4.1.1 Well-founded semantics*

We consider here the well-founded semantics and compare it with the semantics for update programs proposed by Bidoit and Maabout (Bidoit & Maabout, 1997). The two semantics differ in their rewriting of the source program and in the application of the input updates to the source database.

We first recall how programs are rewritten in (Bidoit & Maabout, 1997). Given an update program $\mathcal{UP}$ denote with $st_{[BM]}(\mathcal{UP})$ the standard $\mathtt{DATALOG}^{\neg}$ program obtained from $\mathcal{UP}$ by (1) replacing each rule of the form $+A \leftarrow Body$ (resp., $-A \leftarrow Body$) in $\mathcal{AP}$ with a rule $+A \leftarrow Body, \neg - A$ (resp., $-A \leftarrow Body, \neg + A$) and (2) inserting, for each update atoms $+A \in \delta$ (resp. $-A \in \delta$), a rule of the form $+A \leftarrow \neg - A$ (resp., $-A \leftarrow \neg + A$).

The following example shows the different behavior of the two rewritings.

*Example 8*
Let us consider the update program of Example 2 $\mathcal{UP} = \langle\{+\mathtt{confirm(x,d)}\}, \mathcal{AP}\rangle$ where $\mathcal{AP}$ consists of the rule

$$
\begin{array}{lll}
+\mathtt{mgr(X,D)} & \leftarrow & +\mathtt{confirm(X,D)}\cdot \\
-\mathtt{mgr(X,D)} & \leftarrow & \mathtt{mgr(X,D)},\ \neg + \mathtt{mgr(X,D)},\ +\mathtt{confirm(Y,D)}\cdot
\end{array}
$$

The rewritten program $st(\mathcal{UP})$ consists of the rules

$$
\begin{array}{lll}
+\mathtt{mgr(X,D)} & \leftarrow & \mathtt{confirm}^+\mathtt{(X,D)},\ \neg\mathtt{ck_mgr(X,D)}\cdot \\
-\mathtt{mgr(X,D)} & \leftarrow & \mathtt{mgr(X,D)},\ \neg\mathtt{mgr}^+\mathtt{(x,d)},\ \mathtt{confirm}^+\mathtt{(Y,D)}, \neg\mathtt{ck_mgr(x,d)}\cdot \\
\mathtt{confirm}'\mathtt{(x,d)}\cdot & & \\
\mathtt{ck_mgr(X,D)} & \leftarrow & +\mathtt{mgr(X,D)},\ -\mathtt{mgr(X,D)}\cdot \\
\mathtt{confirm}^+\mathtt{(X,D)} & \leftarrow & +\mathtt{confirm(X,D)}\cdot \\
\mathtt{confirm}^+\mathtt{(X,D)} & \leftarrow & \mathtt{confirm}'\mathtt{(X,D)}\cdot \\
\mathtt{mgr}^+\mathtt{(X,D)} & \leftarrow & +\mathtt{mgr(X,D)}\cdot \\
\mathtt{mgr}^+\mathtt{(X,D)} & \leftarrow & \mathtt{mgr}'\mathtt{(X,D)}\cdot
\end{array}
$$

where the fact $\mathtt{confirm}'\mathtt{(x,d)}$ takes into account the content of $\delta$. The standard program $st_{[BM]}(\mathcal{UP})$ contains the rules

$$
\begin{array}{lll}
+\mathtt{mgr(X,D)} & \leftarrow & +\mathtt{confirm(X,D)},\ \neg - \mathtt{mgr(X,D)}\cdot \\
-\mathtt{mgr(X,D)} & \leftarrow & \mathtt{mgr(X,D)},\ \neg + \mathtt{mgr(X,D)}, +\mathtt{confirm(Y,D)}\cdot \\
+\mathtt{confirm(x,d)} & \leftarrow & \neg - \mathtt{confirm(x,d)}\cdot
\end{array}
$$

Assuming that the source database is empty, the well founded model of $st(\mathcal{UP})_D$ is total and contains as positive atoms $\mathtt{confirm}'\mathtt{(x,d)}, \mathtt{confirm}^+\mathtt{(x,d)}, +\mathtt{mgr(x,d)}$ and $\mathtt{mgr}^+\mathtt{(x,d)}$ whereas the well founded model of $st_{[BM]}(\mathcal{UP})_D$ contains as defined literals only $+\mathtt{confirm(x,d)}$ and $\neg - \mathtt{confirm(x,d)}$ [6] whereas all others literals are undefined. Thus, our semantics correctly concludes that $\mathtt{mgr(x,d)}$ is inserted into the database (as true fact) whereas the semantics proposed in (Bidoit & Maabout, 1997) cannot decide about the insertion of $\mathtt{mgr(x,d)}$ (it inserts the tuple $\mathtt{mgr(x,d)}$ as undefined fact). □

Now let $\mathcal{UP}_{\mathcal{WS}[BM]}(D)$ be the application of $\mathcal{UP}$ to $D$ under the well-founded semantics as proposed in (Bidoit & Maabout, 1997), we have the following result:

*Theorem 1*
$\mathcal{UP}_{\mathcal{WS}[BM]}(D) \preceq \mathcal{UP}_{\mathcal{WS}}(D)$.

**Proof.** See Appendix. □

The main difference between the technique proposed in (Bidoit & Maabout, 1997) and the technique proposed here is in the rewriting of atoms of the input update set. Indeed, given an update program $\langle\delta, \mathcal{AP}\rangle$, the technique proposed in (Bidoit & Maabout, 1997) does not distinguish between atoms in the input update set $\delta$ and the rules in the active program $\mathcal{AP}$ (atoms in $\delta$ are considered as facts of $\mathcal{AP}$) whereas our technique considers the atoms in the input update set $\delta$ as updates performed over the input database which (subsequently) trigger the rules of the active program $\mathcal{AP}$. Moreover, the atoms which are undefined under the semantics proposed in (Bidoit & Maabout, 1997) which are not undefined under our semantics are atoms contained in the input update set $\delta$ and atoms which depend on them. As a consequence, the rewriting proposed here is more intuitive and also leaves fewer atoms undefined.

[6] $-\mathtt{confirm(x,d)}$ is false in all stable models because there is no rule defining it.

### *4.1.2 Max-deterministic semantics*

The well-founded semantics is not able to catch the intuitive meaning of all programs. We show this by means of an example.

*Example 9*
Consider a database $D$ and the update program $\langle \delta, \mathcal{AP} \rangle$ where $\delta = \{+\texttt{new(a)}\}$ and $\mathcal{AP}$ consists of the rules

$$
\begin{array}{lll}
+\texttt{emp(X)} & \leftarrow & \neg + \texttt{mgr(X)},\ +\texttt{new(X)}\cdot \\
+\texttt{mgr(X)} & \leftarrow & \neg + \texttt{emp(X)},\ +\texttt{new(X)}\cdot \\
+\texttt{worker(X)} & \leftarrow & \neg + \texttt{noworker(X)},\ +\texttt{new(X)}\cdot \\
+\texttt{noworker(X)} & \leftarrow & \neg + \texttt{worker(X)},\ +\texttt{new(X)}\cdot \\
+\texttt{worker(X)} & \leftarrow & +\texttt{emp(X)}\cdot \\
+\texttt{worker(X)} & \leftarrow & +\texttt{mgr(X)}\cdot
\end{array}
$$

The rewritten program, where rules with body atoms not defined by any rule have been omitted, since they cannot be used to derive any fact, is

$$
\begin{array}{lll}
+\texttt{emp(X)} & \leftarrow & \neg + \texttt{mgr(X)},\ \texttt{new}'\texttt{(X)},\ \neg\texttt{ck_emp(X)}\cdot \\
+\texttt{mgr(X)} & \leftarrow & \neg + \texttt{emp(X)},\ \texttt{new}'\texttt{(X)},\ \neg\texttt{ck_mgr(X)}\cdot \\
\texttt{new}'\texttt{(X)}\cdot & & \\
+\texttt{worker(X)} & \leftarrow & \neg + \texttt{noworker(X)},\ \texttt{new}'\texttt{(X)},\ \neg\texttt{ck_worker(X)}\cdot \\
+\texttt{noworker(X)} & \leftarrow & \neg + \texttt{worker(X)},\ \texttt{new}'\texttt{(X)},\ \neg\texttt{ck_noworker(X)}\cdot \\
+\texttt{worker(X)} & \leftarrow & +\texttt{emp(X)},\ \neg\texttt{ck_woker(X)}\cdot \\
+\texttt{worker(X)} & \leftarrow & +\texttt{mgr(X)},\ \neg\texttt{ck_worker(X)}\cdot
\end{array}
$$

where the definitions of the predicates `ck_emp`, `ck_mgr`, `ck_worker` and `ck_noworker` have been omitted.

The application of the well-founded semantics gives a model containing as defined literals only `new'(a)` and all other derived atoms are undefined. However, the intuitive meaning of the first two rules is that if `new(a)` is inserted either `emp(a)` or `mgr(a)` should be inserted too whereas the last two rules say that `worker(a)` must be inserted if either `emp(a)` or `mgr(a)` is inserted. The third and fourth rules say that if `new(a)` is inserted either `worker(a)` or `noworker(a)` must be inserted. Therefore, it seems more intuitive to deduce that `worker(a)` should be inserted into the database independently on the particular insertion of `emp(a)` and `mgr(a)`. This 'intuitive' meaning is captured by the max-deterministic semantics which gives, for the rewritten program, a model where the atom +`worker(a)` is true, the atoms +`emp(a)` and +`mgr(a)` are undefined and the atom +`noworker(a)` is false. □

In the above example we have seen a program where the max-deterministic model is more informative than the well-founded model. The next proposition shows that this is generally true.

*Proposition 2*
$\mathcal{UP}_{\mathcal{WS}}(D) \preceq \mathcal{UP}_{\mathcal{MD}}(D)$.

**Proof.** Let $\mathcal{LP} = st(\mathcal{UP})$ and let $M_1$ and $M_2$ be, respectively, the well-founded model and the max-deterministic model of $\mathcal{LP}_D$. Let $D'$ be the database derived from the application of $\delta$ to $D$. Since $\overline{M}_2 \subseteq \overline{M}_1$, from Definition 4 we derive that, $\overline{D}_2 \subseteq \overline{D}_1$, where $D_2 = \hat{M}_2(D')$ and $D_1 = \hat{M}_1(D')$ are, respectively, the databases derived under the max deterministic and the well founded semantics. □

We will see in Section 6 that the advantage of having more informative semantics is counter balanced by the computational complexity.

### *4.1.3 Total deterministic semantics*

In this section we consider a total deterministic semantics for update programs. The idea is to consider as 'correct' only update programs which generate total database and to reject the programs which introduce unknown facts into the database.

*Definition 5*
*Let $D_1$ be a total database and let $\mathcal{UP}$ be an update program. Then, the application of $\mathcal{UP}$ to $D_1$ under total well-founded semantics (resp. max-deterministic semantics), denoted $\mathcal{UP}_{\mathcal{TWS}}(D_1)$ (resp. $\mathcal{UP}_{\mathcal{TMD}}(D_1)$), is equal to $D_2 = \mathcal{UP}_{\mathcal{WS}}(D_1)$ (resp., $D_2 = \mathcal{UP}_{\mathcal{MD}}(D_1)$) if $D_2$ is total, otherwise it is equal to $D_1$.* □

Observe that the condition of totality of the semantics is on the mapping and not on the model of the rewritten program. We next introduce necessary and sufficient conditions which guarantee that the mapping $\mathcal{UP}$ applied to a total database gives a total database.

*Proposition 3*
Let $\mathcal{UP}$ be an update program and let $D$ be a total database. Let $M = SEM_{\mathcal{XS}}[\mathcal{UP}, D]$. Then, $\mathcal{UP}(D)$ is total iff at least one of the following conditions is true

1. $M$ is total, or
2. for each atom $+p \in \bar{M}$ it must be the case that $p \in D^+$, and for each atom $-p \in \bar{M}$ it must be the case that $p \in D^-$.

**Proof.** Let $\mathcal{UP} = \langle \delta, \mathcal{AP} \rangle$ and let $D_1 = \delta(D)$. Clearly $D_1$ is total since $\delta$ is conflict free.

1. If $M$ is total then $\hat{M}(D_1)$ is also total since $\hat{M}$ is conflict free.
2. From Definition 3 we derive that an atom $p \in D_1^+$ (resp. $p \in D_1^-$) is undefined in $\hat{M}(D_1)$ only if $-p$ (resp. $+p$) is undefined in $\hat{M}$. But $\hat{M} \subseteq M$ and, therefore, $\hat{M}(D_1)$ is also total. □

*Example 10*

Consider the update program of Example 9 where the update atoms $+\texttt{emp(X)}$ and $+\texttt{mgr(X)}$ are replaced, respectively, by the standard atoms $\texttt{emp(X)}$ and $\texttt{mgr(X)}$ (clearly, the update literals $\neg + \texttt{emp(X)}$ and $\neg + \texttt{mgr(X)}$ are replaced, respectively, by the literals $\neg\texttt{emp(X)}$ and $\neg\texttt{mgr(X)}$).

Assuming the input database $D$, the application of $\mathcal{UP}$ to $D$ under total well-founded semantics does not modify the database $D$ since the the well-founded model of $st(\mathcal{UP})_D$ contains undefined update atoms, whereas the application of $\mathcal{UP}$ to $D$ under total max-deterministic semantics produces a new database also containing the atoms $\texttt{new(a)}$ and $\texttt{worker(a)}$. □

Observe that the application of the program of Example 3 is not total under the well-founded semantics whereas it is total under the max-deterministic semantics.

#### *4.1.4 Unique total stable model semantics*

We now consider a different semantics which is not based on deterministic models. A Datalog program may have zero, one or more total stable models associated to it. A possible declarative confluent semantics for a program can be given by the unique total stable model semantics. Thus, we give semantics only to programs which have a unique total stable model (programs with zero or more than one total stable models are rejected). It is obvious that this semantics, as well as the previous ones, guarantees termination and confluence. The unique total stable model semantics will be denoted bt $\mathcal{UTS}$.

*Definition 6*

*Let $D_1$ be a total database and let $\mathcal{UP}$ be an update program. Then, the application of $\mathcal{UP}$ to $D_1$ under unique total stable model semantics, denoted $\mathcal{UP}_{\mathcal{UTS}}(D_1)$, is equal to $D_2 = \hat{M}(\delta(D_1))$ if $M$ is the unique total stable model of $st(\mathcal{UP})_{D_1}$, otherwise it is equal to $D_1$.* □

*Example 11*

Consider a database $D$ and the update program $\langle \delta, \mathcal{AP} \rangle$ where $\delta = \{+new(a)\}$ and $\mathcal{AP}$ consists of the rules

$$
\begin{array}{lll}
+\texttt{emp(X)} & \leftarrow & \neg + \texttt{mgr(X)},\ +\texttt{new(X)}\cdot \\
+\texttt{mgr(X)} & \leftarrow & \neg + \texttt{emp(X)},\ +\texttt{new(X)}\cdot \\
+\texttt{worker(X)} & \leftarrow & +\texttt{new(X)}\cdot \\
-\texttt{worker(X)} & \leftarrow & +\texttt{mgr(X)}\cdot
\end{array}
$$

Assuming that the input database $D$ is empty, the program $st(\mathcal{UP})_D$ has a unique total stable model containing the literals $+\texttt{new(a)}$, $+\texttt{emp(a)}$, $\neg{+}\texttt{mgr(a)}$, $+\texttt{worker(a)}$, $\neg - \texttt{worker(a)}$. □

### *4.2 Nonconfluent stable model semantics*

We consider now nonconfluent semantics based on stable models. The presence of more than one stable model does not assure confluence, but in some cases, it captures in a better way the intuitive behavior of active programs. Let us now clarify this by means of an example where all the previously defined semantics fail to handle the update requests returning an incomplete database or rejecting the requests that do not match the intuitive meaning of the program.

*Example 12*
Consider a database $D$ and the update program $\langle \delta, \mathcal{AP} \rangle$ where $\delta = \{+\mathtt{new}(\mathtt{a})\}$ and $\mathcal{AP}$ consists of the rules

$$\begin{array}{lll} +\mathtt{emp}(\mathtt{X}) & \leftarrow & \neg + \mathtt{mgr}(\mathtt{X}),\ +\mathtt{new}(\mathtt{X})\cdot \\ +\mathtt{mgr}(\mathtt{X}) & \leftarrow & \neg + \mathtt{emp}(\mathtt{X}),\ +\mathtt{new}(\mathtt{X})\cdot \\ +\mathtt{worker}(\mathtt{X}) & \leftarrow & +\mathtt{emp}(\mathtt{X})\cdot \\ +\mathtt{worker}(\mathtt{X}) & \leftarrow & +\mathtt{mgr}(\mathtt{X})\cdot \end{array}$$

The first two rules state that if `new(a)` is inserted into the database either `emp(a)` or `mgr(a)` is inserted too. However, in both cases the atom `worker(a)` is inserted into the database. Well-founded and max-deterministic semantics do not capture this behavior and they conclude that the atoms $+\mathtt{emp}(\mathtt{a}), +\mathtt{mgr}(\mathtt{a})$ and `worker(a)` are undefined in the new database. □

To overcome the limits of deterministic semantics we next consider the non-deterministic version of stable model semantics. Non deterministic semantics permits us to have databases without undefined elements but this implies that we renounce the confluence property.

### *4.2.1 Total stable model semantics*

*Definition 7*
`Total stable model semantics.`
*Let $D_1$ be a total database, $\mathcal{UP}$ be an update program and let $S$ be the set of total stable models for $st(\mathcal{UP})_{D_1}$. Then, the application of $\mathcal{UP}$ to $D_1$ under the non-deterministic version of total stable model semantics, denoted $\mathcal{UP}_{\mathcal{TS}}(D_1)$, is equal to (i) $D_2 = \hat{M}(\delta(D_1))$ if $S$ is not empty and $M \in S$ is a (non deterministically selected) model in $S$, or (ii) $D_1$ if $S$ is empty (e.g. there are no total stable model).* □

The problem with this semantics is that in some cases there are programs which do not admit any total stable model, but could admit total transformations under partial stable model semantics (recall that programs admit at least one partial stable model). Although the computation of this semantics is expensive (see below), it guarantees that the transformations are total.

### *4.2.2 Maximal stable model semantics*

To overcome the above mentioned problem, it is possible to consider restricted languages which guarantee the existence of at least one total stable model, computable in polynomial time (Greco *et al.*, 1995). A different approach can be defined by selecting, nondeterministically, a meaningful stable model. We consider maximal stable models which are the natural extension of the total stable model semantics to partial interpretations. The main difference between total and maximal stable models semantics is that the first guarantees totality of the semantics whereas the former guarantees the existence of a model for all programs.

*Definition 8*
`Maximal stable model semantics.`
*Let $D_1$ be a total database, $\mathcal{UP}$ be an update program and let $S$ be the set of maximal stable models for $st(\mathcal{UP})_{D_1}$. Then, the application of $\mathcal{UP}$ to $D_1$ under the non-deterministic version of maximal stable model semantics, denoted $\mathcal{UP}_{\mathcal{MS}}(D_1)$ is equal to $D_2 = \hat{M}(\delta(D_1))$ where $M \in S$ is a (non deterministically selected) model in $S$.* □

This semantics assure that the update transformation can be computed efficiently (see below), but it does not guarantee total transformations and, therefore, total databases. That is, since the maximal model $M$ of $st(\mathcal{UP})_{D_1}$ is chosen in a casual way, $\hat{M}$ may contain undefined update atoms that when applied to $D_1$ return an incomplete database $D_2$.

A further refinement could be obtained by restricting the selection to models which generate total transformations on the database.

*Definition 9*
`Maximal stable model semantics with total transformations.`
*Let $D_1$ be a total database, $\mathcal{UP}$ be an update program and let $S$ be the set of maximal stable models for $st(\mathcal{UP})_{D_1}$. Then, the application of $\mathcal{UP}$ to $D_1$ under the non-deterministic version of maximal stable model semantics with total transformations, denoted $\mathcal{UP}_{\mathcal{MS}_{\mathcal{TT}}}(D_1)$, is equal to (i) $D_2 = \hat{M}(\delta(D_1))$, if $S$ is not empty, $M \in S$ is a non deterministically selected model in $S$ such that $D_2$ is total, or (ii) $D_1$, if $S$ does not contain any model $M$ such that $\hat{M}(\delta(D_1)$ is total.* □

We conclude by noting that the last semantics assures that the selection of the stable model is restricted to those which guarantee that the updated database is total. We will see in Section 5 that the selection of a model which guarantees the totality of the transformation has high computational complexity whereas, by considering also partial transformations, the complexity of the (nondeterministic) selection is polynomial.

## 5 Complexity

As shown in (Abiteboul *et al.*, 1994) most of the operational semantics proposed in the literature have very high complexity and expressivity. The semantics presented in this paper have low complexity and they are all located under the second layer of the polynomial hierarchy. It seems to us that semantics with low complexity are more desirable for database applications.

Classical complexity theory classifies languages on the basis of the difficulty of deciding whether a given input string belongs to the language. I/O (update) programs have been classified in a similar fashion, defining a recognition problem associated to them. However, another equivalent way to define the computational complexity is based on the complexity of constructing the result.

Let $\mathcal{UP}$ be an update program and let $D$ be a database on a fixed schema $\mathcal{S}$. Then, we say that $\mathcal{UP}$, under semantics $\mathcal{XS}$ has complexity $C$, if the complexity of constructing the result for $\mathcal{UP}_{\mathcal{XS}}(D)$ is $C$. The complexity of computing, for a given update program $\mathcal{UP}$ and a database $D$, $\mathcal{UP}(D)$ depends on the underlying semantics, since different semantics have different complexity.

We recall here how basic complexity classes are defined. $\mathcal{P}$ and $\mathcal{NP}$ denote the classes of decision problems computable in polynomial time, respectively, by a deterministic Turing machine and by a nondeterministic Turing machine. $co\mathcal{NP}$ is the class of decision problems whose complementary problems are in $\mathcal{NP}$ and $\mathcal{D}^p$ is the class of decision problems which can be expressed by means of two distinct problems, one in $\mathcal{NP}$ and the second in $co\mathcal{NP}$. $\mathcal{US}$ is the class of decision problems with *unique solution*[7] (Blass & Gurevich, 1982), whereas $\Delta_2^p$ is the class of decision problems computable in polynomial time by a deterministic Turing machine which uses an $\mathcal{NP}$ oracle. It is known that $\mathcal{P} \subseteq \mathcal{NP} \subseteq \mathcal{D}^p \subseteq \Delta_2^p$ and that $\mathcal{P} \subseteq co\mathcal{NP} \subseteq \mathcal{US} \subseteq \mathcal{D}^p \subseteq \Delta_2^p$. Whenever we are interested in finding solutions to problems, we consider complexity classes for search problems (also called function problems). Search and decision problems are closely related and decision problems can be used to state negative complexity results for search problems. (We refer the reader to (Papadimitriou, 1994) for the formal definition of the relationship between decision and search problems.) Here we consider the class $F\mathcal{NP}$, $F\mathcal{US}$ and $F\Delta_2^p$ which are the classes of search problems whose corresponding decision problems belong, respectively, to the classes $\mathcal{NP}$, $\mathcal{US}$ and $\Delta_2^p$ and the class $F\mathcal{P}$ which denotes the class of search problems in $F\mathcal{NP}$ that can be computed in polynomial time.

*Proposition 4*

Let $\mathcal{UP} = \langle \delta, \mathcal{AP} \rangle$ be an update program and let $D$ be a database on a fixed schema $\mathcal{S}$. Then, the complexity of computing $\mathcal{UP}(D)$ is

1. $F\mathcal{P}$, under $\mathcal{WS}$ and non-deterministic $\mathcal{MS}$ semantics;
2. $F\mathcal{NP}$, under non-deterministic $\mathcal{TS}$ and non-deterministic $\mathcal{MS}_{\mathcal{TT}}$ semantics;
3. $\mathcal{D}^p$-hard and in $F\Delta_2^p$, under max-deterministic semantics;
4. $F$-$\mathcal{US}$, under unique total stable model semantics.

**Proof.** First of all observe that the application of $\delta$ to $D$ can be done in polynomial time as well as the rewriting of $\mathcal{UP}$ into $st(\mathcal{UP})$. Moreover, since the cardinality of every minimal model $M$ for $st(\mathcal{UP})_D$ is polynomial in the size of $D$, the application of $M$ to $D' = \delta(D)$ is computable in polynomial time. Therefore we have to consider the complexity of computing the model $M$ of the Datalog$^\neg$ program $st(\mathcal{UP})_D$ under the various semantics.

[7] Formally, $\mathcal{US}$ is the class of languages accepted by a nondeterministic polynomial time bounded Turing machine with the convention that a string is accepted if there is exactly one accepting computation.

1. For a given program $\mathcal{LP}$ and a database $D$, both the well-founded model and any maximal stable model can be computed in polynomial time; thus the complexity of computing $\mathcal{UP}(D)$ under both the well-founded and the non-deterministic version of maximal stable model semantics is $F\mathcal{P}$.
2. The problem of checking if a Datalog$^\neg$ program $\mathcal{LP}$ has a total stable model is $\mathcal{NP}$-complete; clearly the problem of finding a stable model for $st(\mathcal{UP})_D$ is in $F\mathcal{NP}$; therefore, the complexity of computing $\mathcal{UP}(D)$ under non-deterministic total stable model semantics is in $F\mathcal{NP}$.
   Consider now the $\mathcal{MS}_{\mathcal{TT}}$ semantics. Checking if a Datalog program has a maximal stable model $M$ which is total w.r.t. a subset of its predicates is also $\mathcal{NP}$-complete. Therefore, the complexity of computing $\mathcal{UP}(D)$ under non-deterministic $\mathcal{MS}_{\mathcal{TT}}$ semantics is also in $F\mathcal{NP}$.
3. In (Saccà & Zaniolo, 1997) it has been shown that computing the max-deterministic model is $\mathcal{D}^p$ hard and it is in $F\Delta_2^p$; therefore computing $\mathcal{UP}(D)$ under the max-deterministic semantics is $\mathcal{D}^p$-hard and it is in $F$-$\Delta_2^p$.
4. The problem of checking if a Datalog$^\neg$ program has a unique total stable model is in $\mathcal{US}$. Therefore, the complexity of computing $\mathcal{UP}(D)$ under unique total stable model semantics is in $F\mathcal{US}$. □

The complexity of computing update programs under a given semantics is counter balanced by its expressivity. Indeed, update programs are functions on databases and, as shown above, the well-founded semantics only permits the expression of a subset of transformations corresponding to polynomial problems. Total stable model semantics and maximal stable model semantics permit the expression of a larger set of transformations but they have higher complexity. The choice of the right semantics depends on the class of problems which our transformations want express.

## 6 Discussion and Related Work

Active and deductive rules can be seen as opposite ends of a spectrum of database rule languages (Widom, 1993). Deductive rules provide high-level declarative language for specifying intensional relations. In contrast, active rules are more low-level and often need explicit control on rule execution. Most commercial active rule systems operate at a relatively low-level of abstraction and are heavily influenced by implementation-dependent procedural features. The procedural control of active rules makes it very difficult to understand or predict their meaning. This difficulty increases as rules are added or deleted and the meaning of rules is not understood by looking at their specifications. For these reasons, active rules so far have been approached with a great deal of care (Widom & Ceri, 1996). Several tools have been provided to understand the run-time behaviour of active rules. These tools have been developed for compile-time rule analysis to understand if rule processing terminates and to know if the execution of a set of rules which are not fully prioritized always gives a unique outcome.

Several techniques, based on syntactic and semantics analysis of rules, have been proposed to detect termination. Syntactic analysis is based on the construction of a *triggering graph* (Aiken *et al.*, 1994). The nodes of the graph are rules whereas there is an arc from node $r_i$ to node $r_j$ if the execution of rule $r_i$ can trigger rule $r_j$. If there is a cycle in the graph, the rules may trigger each other indefinitely and the termination property is not guaranteed. This idea was refined in (Baralis & Widom, 1994) where a different graph, called *activation graph*, was proposed and in (Baralis *et al.*, 1996) where a combination of triggering and activation graphs was proposed. A different technique based on abstract interpretation has recently been proposed in (Bailey *et al.*, 1997). It is worth noting that all the proposed techniques can check termination for restricted cases, i.e. they are able to determine possible cases of nontermination but they are not able to determine termination in the general case. However, generally, checking whether an arbitrary set of rules terminates is an undecidable problem. The problem remains undecidable also for rules which do not use arithmetics and function symbols.

Confluence is guaranteed if rules are prioritized, i.e. if there is a linear order on priorities associated with rules, so that only one rule at time can be activated. Moreover, confluence can also be guaranteed in cases where more than one rule can be activated at the same time. Confluence tools are based on the *commutativity* of rules execution. Two rules $r_i$ and $r_j$ commute if, for each database $D$, the activation of $r_i$ followed by $r_j$ and the activation of $r_j$ followed by $r_i$, produce the same database. A basic technique for checking confluence is based on the checking of commutativity for each pair of rules in the rule set. Less restrictive conditions have been defined in the literature. These are based on finding minimal sets of rules which must be commutative (Aiken *et al.*, 1994). However, generally, checking confluence for an arbitrary set of rules is undecidable.

The analysis of rules can be further complicated by aspects such as immediate versus deferred triggering, the use of delta relations and composite events. Although there has been considerable research and development in the area of active databases, there has been very little activity in the study of formal foundations. Some preliminary work, based on the rewriting of active rules in terms of deductive rules, can be found in (Baral & Lobo, 1996; Zaniolo, 1993), whereas in (Widom, 1992) the semantics of active rules is given in terms of denotational semantics. The combination of deductive and active rules has been also investigated (Lausen *et al.*, 1998; Ludascher, 1998; Zaniolo, 1993). All these approaches are based on the simulation of active rules by means of deductive rules, where atoms have associated a temporal argument (also called state argument).

The approach proposed in this paper is based on the definition of a declarative semantics for active (and deductive) rules. Our semantics guarantees i) termination, for programs without function symbols and ii) confluence, under a confluent semantics.

In this paper we have not investigated the presence of arithmetical operators and function symbols. However, since the semantics we have consider apply to general logic programs, confluent semantics guarantee confluence in the general case.

The presence of arithmetical operators and function symbols does not permit to guarantee termination since the Herbrand base is infinite. A large amount of work has been devoted to the termination of query evaluation in logic programming and deductive databases (De Schreye & Decorte, 1994; Kifer *et al.*, 1988; Ramakrishnan *et al.*, 1987; Krishbamurty *et al.*, 1988; Sagiv & Vardi, 1989). Distinction is made between the problem of proving termination under top-down evaluation and proving that bottom-up evaluation generates finite answers. The latter property is known as *safety*. For termination under top-down evaluation we refer to the survey of De Schreye and Decorte (De Schreye & Decorte, 1994). The safety property for queries with function symbols, under bottom-up evaluation, has been analyzed in (Kifer *et al.*, 1988; Ramakrishnan *et al.*, 1987; Krishbamurty *et al.*, 1988; Sagiv & Vardi, 1989). A related problem is whether the evaluation strategy computes all answers and terminates, i.e. all intermediate relations are finite. This problem is also known as *termination* or *effective computability*. Clearly, effective computability is a stronger property than safety, i.e. effective computability implies safety but the converse is not true. The problem of determining safety for queries with function symbols is, generally, undecidable.

Several techniques for checking the effective computability of special classes of queries have been proposed in the literature. Most of the proposed techniques are based on the approximation of queries with function symbols by means of queries without function symbols over infinite relations. We mention here the restricted cases known as *supersafety* problem (i.e. the question of whether a query has a finite answer in any model which is a fixpoint of the operator $T_{\mathcal{LP}}$ (Lloyd, 1987)), which has been shown to be decidable and axiomatizable (Kifer *et al.*, 1988) and *strong safety* problem (i.e. the question of whether programs, where derived predicates used to compute other derived predicates are safe, are effectively computable), which has been shown to be decidable and efficiently computable. Another class of interest which has been studied is that of monadic queries, i.e. queries in which all intensional predicates are monadic. It has been shown that, for monadic queries effective computability is decidable in polynomial time (Sagiv & Vardi, 1989). Other techniques based on the identification of cycles performing monotonic operations with a finite bound have also been studied.

## 7 Conclusions

In this paper we have analyzed declarative semantics for active rules. The advantages of using declarative semantics are that termination is guaranteed for function free queries, confluence is also guaranteed by using confluent semantics and the semantics is well defined and intuitive. Furthermore, the framework proposed in this paper permits a natural integration of deductive and active rules and can also be applied to queries with function symbols or to queries over infinite databases. We have also shown that semantics which reduce the set of unknown literals have high complexity (and also high expressive power). Further work should be devoted to investigating the relation between non-deterministic and procedural semantics; in both cases different computations of the same program on the same database could result in different outcomes.

*Acknowledgments.* Work partially supported by the EU projects “Telcal” and “Contact” and by the Murst Project “Interdata”. The authors would like to thank the anonymous referees for their helpful comments on the original version of the paper.

## References

Abiteboul S., Hull R., & Vianu, V. (1994) *Foundations of Databases*, Addison-Wesley.

Abiteboul, S., Simon, E. & Vianu, V. (1990) Non-deterministic languages to express deterministic transformations, In *Proceedings ACM Symp. on Principles of Database Systems*, 218–229.

Abiteboul, S. & Vianu, V. (1991) Datalog Extensions for Databases Queries and Updates. In *Journal of Computer and System Sciences*, **43**(1), 62–124.

Aiken, A., Widom, J. and Hellerstein, J. M. (1992) Behaviour of database production rules: termination, confluence and observable determinism. In *Proceedings ACM SIGMOD International Conference on Management of Data*, 59–68.

Bailey, J., Crnogorac, L., Ramamoharanao, K. & Sondergaard, H. (1997) Abstract interpretation of active rules and its use in termination analysis. In *Proceedings International Conference on Database Theory*, 188–202.

Baralis, E., Ceri, S. & Paraboschi, S. (1996) Improved rule analysis by means of triggering and activation graphs. In T. Sellis (ed.) *Rules in Database Systems*, 165–181. Springer-Verlag.

Baralis, E. & Widom, J. (1994) An algebraic approach to rule analysis in expert database systems. In *Proceedings International Conference on Very Large Data Bases*, 475–486.

Baral, C. & Lobo, J. (1996) Formal characterization of active databases. In *Proceedings International Workshop on Logic in Databases*, 175–195.

Bidoit, N. & Maabout, S. (1997) A Model Theoretic Approach to Update Rule Programs. In *Proceedings International Conference on Database Theory*, 173–187.

Blass, A. & Gurevich, Y. (1982) On the Unique Satisfiability Problem. *Information and Control*, 80-88.

Ceri, S. & Fraternali, P. (1997) *The IDEA Methodology*, Addison Wesley.

De Schreye, D. & Decorte, S. (1994) Termination of logic programs: the never-ending story. *Journal of Logic Programming*, 19/20, 199–260.

Gelfond, M., & Lifschitz, V. (1988) The Stable Model Semantics for Logic Programming. In *Proceedings 5th International Conference on Logic Programming*, 1070–1080.

Greco, S. & Saccà, D. (1996) ‘Possible is Certain’ is desirable and can be expressive. *Annals of Mathematics and Artificial Intelligence*, **19** (1-2) 147–168.

Greco, S. & Saccà, D. (1997) Complexity and Expressive Power of Deterministic Stable Model Semantics. In *Proceedings International Conference on Deductive and Object Oriented Databases*, 337–350.

Greco, S., Saccà, D. & Zaniolo, C. (1995) DATALOG Queries with Stratified Negation and Choice: from $\mathcal{P}$ to $\mathcal{D}^p$, In *Proceedings of the International Conference on Database Theory*, 82–96.

Grumbach, S., Lacroix, Z. & Lindell, S. (1995) Implicit Definitions on Finite Structures, In *Proceedings of the Conference on Computer Science Logic*, 252–265.

Johnson, D.S. (1990) A Catalog of Complexity Classes, in J. van Leewen (ed.), *Handbook of Theoretical Computer Science*, Vol. 1, pp. 67–161. North-Holland.

Kifer, M., Ramakrishnan, R. & Silbershatz, A. (1988) An axiomatic approach to deciding query safety in deductive databases. In *Proceedings ACM Symposium on Principles of Database Systems*, pp. 52–60.

Kolaitis, P. G. (1990) Implicit definability of finite structures and unambiguous computations. In *Proceedings 5th IEEE Symposium on Logic in Computer Science*, 168–180.

Kolaitis, P. G. & Papadimitriou, C. (1991) Why not negation by fixpoint. *Journal of Computer and System Sciences*, **43** (1), 125–144.

Krishbamurty, R., Ramakrishnan, R. & Shmueli, O. (1988) A framework for testing safety and effective computability of extended datalog. In *Proceedings ACM SIGMOD Conference on Management of Data*, 154–163.

Lausen, G., Ludascher, B. & May, W. (1998) On Logical Foundations of Active Databases. In *Logics for Databases and Information Systems*, 389-422

Ludascher, B. (1998) *Integration of Active and Deductive Database Rules.* Phd Thesis, Institut für Informatik, Universität Freiburg, Germany.

Lloyd, J.W. (1987) *Foundations of Logic Programming*, Springer-Verlag, Berlin.

Marek, W. & Truszczynski, M. (1991) Autoepistemic Logic. In *Journal of ACM*, **38** (3), pp. 588–619.

Papadimitriou, C. (1994) *Computational Complexity*, Addison-Wesley.

Picouet, P. & V. Vianu, V. (1995) Semantics and expressiveness issues in Active Databases. *Proceedings International Symposium on Principles of Database Systems*, 126–138.

Przymusinski, T.C. (1990) Well-founded Semantics Coincides with Three-valued Stable Semantics, *Foundamenta Informaticae*, **13**, 445–463.

Ramakrishnan, R., Bancilhon, F. & A. Silbershatz, A. (1987) Safety of recursive horn clauses with infinite relations. In *Proceedings ACM Symposium on Principles of Database Systems*, 328–339.

Saccà D. (1997) The Expressive Powers of Stable Models for Bound and Unbound Queries, *Journal of Computer and System Sciences*, **54** (3), 441-464.

Saccà, D. & Zaniolo, C. (1990) Stable Models and Non-Determinism in Logic Programs with Negation, In *Proceedings ACM Symposium on Principles of Database Systems*, 205–218.

Saccà, D. & Zaniolo, C. (1997) Deterministic and Non-Deterministic Stable Models, *Journal of Logic and Computation*, **7** (5), 555–579.

Sagiv, Y. & Vardi, M. (1989) Safety of Datalog queries over infinite databases. In *Proceedings ACM Symposium on Principles of Database Systems*, 160–171.

Ullman, J.D. (1988) *Principles of Database and Knowledge Base Systems*, Computer Science Press.

Van Gelder, A., Ross, K. & Schlipf, J. S. (1991) The Well-Founded Semantics for General Logic Programs, *Journal of the ACM*, **38** (3), 620–650.

Vardi M.Y. (1982) The Complexity of Relational Query Languages. In *Proceedings ACM Symposium on Theory of Computing*, 137–146.

Widom, J. (1992) A denotational semantics for starbust production rule language. In *SIGMOD Record*, **21** (3), 4–9.

Widom, J. (1993) Deductive and Active Databases: two paradigms or ends of a spectrum. In *Proceedings International Workshop on Rules in Database Systems*, 306–315.

Widom, J. & Ceri, S. (eds.) (1996) *Active Databases: Triggers and rules for Advanced Database Processing*, Morgan-Kaufman.

Zaniolo, C. (1993) A unified semantics for Active and Databases. In *Proceedings International Workshop on Rules in Database Systems*, 271–287.

Zaniolo, C. (1995) Active Database Rules with Transaction-Conscious Stable-Model Semantics. In *Proceedings International Conference on Deductive and Object Oriented Databases*, 55–72.

## Appendix

In this appendix we recall the definition of operator $\mathbf{W}_{\mathcal{LP}}$ used to introduce the well-founded semantics and report the proof of Theorem 1.

*Definition 10*

Well Founded model. *Let $\mathcal{LP}$ be a Datalog program and let $X \subseteq (B_{\mathcal{LP}} \cup \neg B_{\mathcal{LP}})$ be an interpretation. The transformation $\mathbf{W}_{\mathcal{LP}}(X)$ is equal to $\mathbf{T}_{\mathcal{LP}}(X) \cup \neg \mathbf{U}_{\mathcal{LP}}(X)$, where $\mathbf{T}_{\mathcal{LP}}$ is the immediate consequence operator and $\mathbf{U}_{\mathcal{LP}}$ is the greatest unfounded w.r.t. X as explained in Section 2.*

*Furthermore, let $\alpha$ be a successor ordinal, we recursively define $\mathbf{W}^{\alpha}_{\mathcal{LP}}(X)$ as X if $\alpha = 0$ and as $\mathbf{W}_{\mathcal{LP}}(\mathbf{W}^{\alpha-1}_{\mathcal{LP}}(X))$ otherwise.*

*Any fixpoint of this transformation represents a P-stable model of $\mathcal{LP}$ and particularly the least fixpoint of $\mathbf{W}_{\mathcal{LP}}$ is the well-founded model of $\mathcal{LP}$.* □

**Theorem 1** $\mathcal{UP}_{\mathcal{WS}[BM]}(D) \preceq \mathcal{UP}_{\mathcal{WS}}(D)$

**Proof.** Let $P_1 = st_{[BM]}(\mathcal{UP})_D$ and $P_2 = st(\mathcal{UP})_D$ and let $M_1$ and $M_2$ be the well-founded models, respectively, of $P_1$ and $P_2$. We must show that $\hat{M}_1(D) \preceq \hat{M}_2(\delta(D))$. Thus, it is sufficient to show that $M_1^+ \subseteq M_2^+ \cup K$ and $M_1^- \subseteq M_2^-$ where $K = \{+p(t) | +p(t) \in \delta \wedge \neg - p(t) \in M_2^-\} \cup \{-p(t) | -p(t) \in \delta \wedge \neg + p(t) \in M_2^-\}$, i.e., $K$ is the set of updates in $\delta$ which are not contradicted in $M_2$.

Table 1 shows the relationship between the programs $P_1$ and $P_2$.

| $\delta$ | $\mathcal{AP}$ | $st_{[BM]}(\mathcal{UP})$ | $st(\mathcal{UP})$ |
|---|---|---|---|
| | $\pm p(t) \leftarrow Body\cdot$ | $\pm p(t) \leftarrow Body, \neg \mp p(t)\cdot$ | $\pm p(t) \leftarrow Body_1, \neg ck_p(t)\cdot$ |
| | $p(t) \leftarrow Body\cdot$ | $p(t) \leftarrow Body\cdot$ | $p(t) \leftarrow Body_1\cdot$ |
| +p(t) | | $+p(t) \leftarrow \neg - p(t)\cdot$ | $p'(t) \leftarrow \cdot$ |
| -p(t) | | $-p(t) \leftarrow \neg + p(t)\cdot$ | $p''(t) \leftarrow \cdot$ |

Table 1. Relations among rules of $st_{[BM]}(\mathcal{UP})$ and $st(\mathcal{UP})$

Each row of the table shows how atoms in $\delta$ (first column) and rules in the active program (second column) are rewritten by the technique proposed in (Bidoit & Maabout, 1997) (third column) and by our technique (fourth column). Observe that *Body* denotes a conjunction of literals (update and standard) whereas $Body_1$ is derived from *Body* but all update atoms are replaced by disjunction (see Section 4). Note the ground instantiations of the rewritten programs. Since auxiliary atoms are defined by facts, disjunctions in $Body_1$ are eliminated by replacing the auxiliary atoms with their truth value. Consequently, auxiliary facts can also be eliminated from $ground(P_2)$.

Therefore, for each rule $r_2$ in $ground(P_2)$ there is a rule $r_1$ in $ground(P_1)$ such that $Body_1 \subseteq Body$.

Now we can show the above containment by induction on the fixpoint operator $\mathbf{W}_P$.

*Step 0.* $I_1 = I_2 = \emptyset$.

*Step* $i > 0$.

Consider two interpretations $I_1$ and $I_2$ for $P_1$ and $P_2$, respectively, such that $I_1 \subseteq (I_2 \cup K_{I_2})$. We must show that $\mathbf{W}(I_1) \subseteq (\mathbf{W}(I_2) \cup K_{\mathbf{W}(I_2)})$.

Consider, first, the containment $\mathbf{T}_{P_1}(I_1) \subseteq \mathbf{T}_{P_2}(I_2) \cup K_{\mathbf{W}(I_2)}$. For each ground predicate $a(t) \in I_1$ there are two possibilities:

1. $[a(t) \notin \delta\cdot]$ This means that there is a rule of the form $\pm a(t) \leftarrow Body, \neg \mp a(t)$ (resp., $a(t) \leftarrow Body$ if $a(t)$ is not an update atom) in $ground(P_1)$ and that $(Body \cup \neg \mp a(t)) \subseteq I_1$. Moreover, there must be a rule $\pm a(t) \leftarrow Body_1, \neg ck_a(t)$ (resp., $a(t) \leftarrow Body_1$ if $a(t)$ is not an update atom) in $ground(P_2)$ and since (1) $Body_1 \subseteq Body \subseteq I_1$, (2) $Body_1$ does not contain input update literals and $I_1 \subseteq I_2 \cup K_{I_2}$ we conclude that $Body_1 \subseteq I_2$. Furthermore, $\neg ck_a(t) \in I_2$ if either $\neg + a(t)$ or $\neg - a(t)$ is in $I_1$.
2. $[a(t) \in \delta\cdot]$ If an update atom $a(t) = +p(t)$ (resp., $a(t) = -p(t)$) belongs to $I_1$ then also $\neg - p(t)$ (resp., $\neg + p(t)$) belongs to it. But since $I_1^- \subseteq I_2^-$, from the definition of $K_I$, we derive that $a(t) \in K_{I_2}$.

Consider, now the relation $\mathbf{U}_{P_1}(I_1) \subseteq \mathbf{U}_{P_2}(I_2)$. Let a(t) be a ground atom belonging to $\mathbf{U}_{P_1}(I_1)$; for each rule $r_2$ of $ground(P_2)$ with $H(r_2) = a(t)$ there is a corresponding rule $r_1$ in $ground(P_1)$ with $H(r_1) = a(t)$. We have that

1. If $(B(r_1) \cap \neg I_1) \neq \emptyset$ (the body of $r_1$ is false w.r.t. $I_1$), then $(B(r_2) \cap \neg I_2) \neq \emptyset$ since the input update cannot be assumed false in $I_1$.
2. If $(B(r_1) \cap \mathbf{U}_1) = \mathbf{U}_{r_1} \neq \emptyset$, then, every atom $a(t) \in \mathbf{U}_{r_1}$ cannot be an input update, otherwise it will be assumed false in $\mathbf{W}_{P_1}(I_1)$, and, therefore, it must appear also in $B(r_2)$.

So we can assume that all the element of $\mathbf{U}_{P_1}(I_1)$ must also belong to $\mathbf{U}_{P_2}(I_2)$.

Finally, $ck_a(t) \in \mathbf{U}_{P_2}(I_2)$ if either $+a(t)$ or $-a(t)$ is in $\mathbf{U}_{P_1}(I_1)$. □